\pdfoutput=1

\documentclass[11pt,a4paper]{article}
\usepackage[hyperref]{acl2021}

\usepackage{times}
\usepackage{latexsym}
\aclfinalcopy 
\def\aclpaperid{74}

\usepackage[T1]{fontenc}

\usepackage[utf8]{inputenc}

\usepackage{microtype}

\usepackage{graphicx}
\usepackage{amssymb}
\usepackage{amsmath}

\title{The Curse of Dense Low-Dimensional Information Retrieval for Large Index Sizes}

\author{Nils Reimers and Iryna Gurevych \\
	Ubiquitous Knowledge Processing Lab (UKP-TUDA)\\
	Department of Computer Science, Technical University of Darmstadt\\
	\url{www.ukp.tu-darmstadt.de}}

\begin{document}
\maketitle
\begin{abstract}
Information Retrieval using dense low-dimensional representations recently became popular and showed out-performance to traditional sparse-representations like BM25. However, no previous work investigated how dense representations perform with large index sizes. We show theoretically and empirically that the performance for dense representations decreases quicker than sparse representations for increasing index sizes. In extreme cases, this can even lead to a tipping point where at a certain index size sparse representations outperform dense representations. We show that this behavior is tightly connected to the number of dimensions of the representations: The lower the dimension, the higher the chance for false positives, i.e.\ returning irrelevant documents.
\end{abstract}

\section{Introduction}
Information retrieval traditionally used sparse representations like TF-IDF or BM25 to retrieve relevant documents for a given query. However, these approaches suffer from the lexical gap problem \cite{berger_lexical_gap}.

To overcome this issue, dense representations have been proposed \cite{end-to-end-retrieval}: Queries and documents are mapped to a dense vector space and relevant documents are retrieved e.g.\ by using cosine-similarity. Out-performance over sparse lexical approaches has been shown for various datasets \cite{end-to-end-retrieval,guo2020multireqa,guu2020realm,gao2020complementing}.

Previous work showed the out-performance for fixed, rather small indexes. The largest dataset where it has been shown is the MS Marco \cite{msmarco} passage retrieval dataset, where retrieval is done over an index of 8.8 million text passages. However, in production scenarios, index sizes quickly reach 100 millions of documents.

We show in this paper, that the performance for dense representations can decrease quicker for increasing index sizes than for sparse representations. For a small index of e.g.\ 100k documents, a dense approach might clearly outperform sparse approaches. However, with a larger index of several million documents, the sparse approach can outperform the dense approach.

We show theoretically and empirically that this effect is closely linked to the number of dimensions for the representations: Using fewer dimensions increases the chances for false positives. This effect becomes more severe with increasing index sizes.

\section{Related Work}
A common choice for dense retrieval is to fine-tune a transformer network like BERT \cite{devlin2018bert} on a given training corpus with queries and relevant documents \cite{guo2020multireqa,guu2020realm,gao2020complementing,dpr,luan2020sparse}. Recent work showed that combining dense approaches with sparse, lexical approaches can further boost the performance \cite{luan2020sparse,gao2020complementing}.  While the approaches have been tested on various information and question answering retrieval datasets, the performance was only evaluated on fixed, rather small indexes. \newcite{guo2020multireqa} evaluated approaches for eight different datasets having index sizes between 3k and 454k documents. 

We are not aware of previous work that compares sparse and dense approaches for increasing index sizes and the connection to the dimensionality. The only work we are aware of that systematically studies the encoding size for dense approaches is \cite{luan2020sparse}, but they only studied the connection to the document length.

\section{Theory} \label{sec_theory}
Dense retrieval approaches map queries and documents\footnote{We use \textit{document} as a cover-term for text of any length.} to a fixed size dense vector. The most relevant documents for a given query can then be found using cosine-similarity.

Using as few dimensions as possible is desirable, as it decreases the memory requirement to store (an index) of millions of vectors and leads to faster retrieval. However, as we show, lower-dimensional representations can have issues with large indices.

Given a query vector $q \in \mathbb{R}^k$, we search our index of document vectors $d_1, ..., d_n \in \mathbb{R}^k$ for the documents that maximizes:
$$\text{cossim}(q, d_i) = \text{cos}(\theta) = \frac{q \cdot d_i}{\left\| q \right\| \left\| d_i \right\|}$$

Note: In the following we just show the case for cosine similarity. The proof extends to other similarity functions like dot-product and any p-norm (Manhatten, Euclidean) as long as the vector space is finite. A finite $n$-dimensional vector space can be mapped to an $n+1$-dimensional vectors space with vectors of unit length. In that case, dot-product in $n$ dimensions is equivalent to cosine-similarity in $n+1$ dimensions. Similar, any $p$-norm in $n$ dimensions can be re-written as cosine-similarity in $n+1$ dimensions.

\textbf{Theorem:} The probability for false positives (I) increases with the index size $n$ and (II) with the decreasing dimensionality $k$.  

\textbf{Proof (I):} Given a query $q$ and the relevant document $d_r$. For simplicity, we assume only a single relevant document. If multiple documents are relevant, we consider only the one with the highest cosine similarity. In order that no false positive is returned, $\text{cossim}(q, d_{r})$ must be greater than $\text{cossim}(q, d_i)$ for all $i \neq r$. Assume the possible vectors are independent. Then, the probability for a false positive is 
$$P(\text{false positive}) = 1 - (1-P(\text{false positive}_i))^{n-1}$$
for an index with $n-1$ negative elements and $P(\text{false positive}_i)$ the probability that a single element is a false positive, i.e.\ $\text{cossim}(q, d_i)  > \text{cossim}(q, d_{r})$. 
 
\textbf{Proof (II):} While the previous proof is straightforward, that the chance of false positives increases with larger index sizes, the more interesting aspect is the relation to the dimensionality, i.e., what is the probability $P(\text{false positive}_i)$ = $P(\text{cossim}(q, d_i) > \text{cossim}(q, d_{r}))$ for a random $d_i$? We show that this probability decreases with more dimensions. 

Without loss of generality, we assume that the vectors are of unit length. The vectors are then on an $k$-dimensional sphere with radius 1. A false positive happens if $\text{cossim}(q, d_i) > \text{cossim}(q, d_{r})$, or, equivalent if $1-\text{cossim}(q, d_i) < 1 - \text{cossim}(q, d_{r})$. I.e., we intersect the  sphere in $k$ dimensions with a hyperplane in $k-1$ dimensions. The area of the cut-off portion is defined by $1-\text{cossim}(q, d_{r})$.  All vectors within the cut-off portion (i.e.\ spherical cap) are false positives. The probability that a random vector will be returned as false positive is:

$$P(\text{false positive}_i)  = A_{cap} / A_{sphere}$$

with $A_{cap}$ the surface area of the spherical cap and $A_{sphere}$ the surface area of the sphere in $k$ dimensions. Define the surface area of the sphere in $k$ dimensions as $A_k$, then the surface area of $A_{cap}$ is \cite{Li2011ConciseFF}:

$$A_{cap} = \frac{1}{2} A_k I_{sin^2 \theta}\left(\frac{k-1}{2}, \frac{1}{2}\right)$$

with $I_x(a,b)$ the regularized incomplete beta function and $\theta$ the polar angle, i.e.\ the angle between $q$ and the relevant document $d_r$. Hence:
 
\begin{gather} \label{eq_false_pos}
P(\text{false positive}_i)  = \frac{1}{2} I_{sin^2 \theta}\left(\frac{k-1}{2}, \frac{1}{2}\right)
\end{gather}

For constant cosine similarity between query $q$ and relevant document $d_r$, $I_{sin^2 \theta}\left(\frac{k-1}{2}, \frac{1}{2}\right)$ is a monotonically decreasing function with increasing dimension $k$. In conclusion, more dimensions decrease the probability for false positives.

Combining (I) and (II) shows that a low dimensional representation might work well for small index sizes. However, with more indexed documents, the probability of false positives increases faster for low dimensional representations than for higher dimensional representations. Hence, at some index size, higher dimensional representations might outperform the lower-dimensional representation.

\section{Empirical Investigation}
In the proof, we have assumed that vectors are independent and uniformly distributed over the space, which gives us a lower bound on the false positive rate. However, in practice, dense representations are neither independent nor uniformly distributed. As shown in \cite{ethayarajh-2019-contextual,li-etal-2020-sentence}, dense representations derived from pre-trained Transformers like BERT map to an anisotropic space, i.e., the vectors occupy only a narrow cone in the vector space. This drastically increases the chance that an irrelevant document is closer to the query embedding than the relevant document. Hence, we study how actual dense models are impacted by increasing index sizes and lower-dimensional representations.

\subsection{Dataset}
We conduct our experiments on the MS MARCO passage dataset \cite{msmarco}. It consists of over 1 million unique real queries from the Bing search engine, together with 8.8 million paragraphs from heterogeneous web sources. Most of the queries have only 1 passage judged as relevant, even though more can exist. The development set consists of 6980 queries and the performance is evaluated using mean reciprocal rank MRR@10. 

To better compare the relative performance differences, we compute 
 a rank-aware error rate:
$$\text{Err} = \frac{1}{n} \sum_{i=1}^n \left(1 - \frac{1}{\text{rank}_i}\right)$$
with $\text{rank}_i$ being the rank of the relevant document for the $i$-th query. To be compatible with MRR@10, we set $\text{rank}_i = \infty$ for $\text{rank}_i > 10$. We then define the relative error rate as $\text{Err}_{\text{Dense}} / \text{Err}_{\text{BM25}}$. A relative error rate of $50\%$ indicates that the dense approach makes only 50\% of the errors compared to BM25 retrieval.

\subsection{Model}
For sparse, lexical retrieval, we use ElasticSearch, which is based on BM25. For dense retrieval, we use a DistilRoBERTa-base model \cite{sanh2020distilbert} as a bi-encoder: The query and the passage are passed independently to the transformer model and the output is averaged to create fixed-sized representations. We train this using InfoNCE loss \cite{InfoNCE}:
$$L = -\log \frac{\exp(\tau \cdot \text{cossim}(q, p_+))}{\sum_i \exp(\tau \cdot \text{cossim}(q, p_i))} $$

with $q$ the query, $p_+$ the relevant passage. We use in-batch negative sampling and use the other passages in a batch as negative examples. We found that $\tau=20$ performs well. We train the model in two setups: 1) only with random (in-batch) negatives, and 2) we provide for each query additionally one hard-negative passage. We use the hard-negative passages provided by the MS MARCO dataset, which were retrieved using lexical search. Models are trained with a batch size of 128 with Adam optimizer and a learning rate of $2e-5$. 

DistilRoBERTa produces representations with 768 dimensions. We also experiment with lower-dimensional representations. There, we added a linear projection layer on-top of the mean pooling operation to down-project the representation to either 128 or 256 dimensions. Dense retrieval is performed using cosine similarity with exact search.

Models were trained using the SBERT framework \cite{reimers-2019-sentence-bert}.\footnote{\url{https://www.SBERT.net}}

\section{Experiments}
First, we study the impact of increasing index sizes with real text passages. Then, we study the performance when random noise is added.
	
\subsection{Increasing Index Size}
In the first experiment, we start with an index that only contains the 7433 relevant passages for the 6980 queries. Then, we add step-wise randomly selected passages from the MS MARCO corpus to the index until all 8.8 million passages are indexed.

\begin{table}[h]
	\centering 
	\footnotesize
	\begin{tabular}{|l|c|c|c|c|}
		\hline
		\textbf{Model} &  \textbf{10k}  & \textbf{100k} & \textbf{1M} & \textbf{8.8M}  \\ \hline
		BM25 & 79.93 & 63.88 & 40.14 & 17.56  \\ \hline
		\multicolumn{5}{|l|}{Trained without hard negatives} \\ \hline
		\quad 128 dim & 87.50 & 68.63 & 39.76  & 15.71  \\
		\quad 256 dim & 88.82 & 70.79 & 41.74 & 17.08 \\ 
		\quad 768 dim & 88.99 & 71.06 & 42.24 & 17.34  \\ \hline
		\multicolumn{5}{|l|}{Trained with hard negatives} \\ \hline
		\quad 128 dim & 90.32 & 77.92 & 54.45 & 27.34 \\
		\quad 256 dim & 91.10 & 78.90 & 55.51 & 28.16 \\ 
		\quad 768 dim & 91.48 & 79.42 & 56.05 & 28.55 \\ \hline
	\end{tabular}
	\caption{Dev performance (MRR@10 $\times 100$) on MS MARCO passage dataset with different index sizes. Higher score = better.}
	\label{table_increase}
\end{table}

\begin{table}[h]
	\centering 
	\footnotesize
	\begin{tabular}{|l|c|c|c|c|}
		\hline
		\textbf{Model} &  \textbf{10k}  & \textbf{100k} & \textbf{1M} & \textbf{8.8M}  \\ \hline
		\multicolumn{5}{|l|}{Trained without hard negatives} \\ \hline
		\quad 128 dim & 62.3 & 86.8 & 100.6  & 102.2  \\
		\quad 256 dim & 55.7 & 80.9 & 97.3 & 100.6 \\ 
		\quad 768 dim & 54.9 & 80.1 & 96.5 &  100.3  \\ \hline
		\multicolumn{5}{|l|}{Trained with hard negatives} \\ \hline
		\quad 128 dim & 48.2 & 61.1 & 76.1 & 88.1 \\
		\quad 256 dim & 44.3 & 58.4 & 74.3 & 87.1 \\ 
		\quad 768 dim & 42.5 & 57.0 & 73.4 & 86.7 \\ \hline
	\end{tabular}
	\caption{Relative error rate (\%) of dense approaches in comparison to BM25 retrieval. Lower score = better.}
	\label{table_increase_err}
\end{table}

Table \ref{table_increase} shows the MRR@10 performance for the different systems. Increasing the index naturally decreases the performance for all systems, as retrieving the correct passages from a larger index is  more challenging. The dense approach trained without hard negatives clearly outperforms BM25 for an index with 10k - 1M entries, but with all 8.8 million passages it performs worse than BM25.

Table \ref{table_increase_err} shows the relative error rate in comparison to BM25 retrieval. For small index sizes, we observe that dense approaches drastically reduce the error rate compared to BM25 retrieval. With increasing index sizes, the gap closes.

\subsection{Index with Random Noise}

MS MARCO is sparsely labeled, i.e., there is usually only a single passage labeled as relevant even though multiple passages would be considered as relevant by humans \cite{trec-dl-2019}. To avoid that the drop in performance is due to the retrieval of relevant, but unlabeled passages, we perform an experiment where we add random irrelevant noise to the index. Our index consists only of the relevant passages and a large fraction of irrelevant, randomly generated strings.\footnote{Strings are generated randomly using lowercase characters a-z and space.} 

We also evaluate the popular DPR system by \newcite{dpr}, which is a BERT-based dense retriever trained on the Natural Questions (NQ) dataset \cite{NQ_dataset}. We chose the NQ dev set, consisting of 1772 questions from Google search logs. DPR encodes the passage as \texttt{Title [SEP] Paragraph}. We create a random string for the paragraph and combine it with 1) a randomly generated string as title, 2) selecting randomly one of the over 6 Million real Wikipedia article titles, 3) selecting randomly one of the 1772 article titles found in the NQ dev set.

\begin{table}[h]
	\centering 
	\footnotesize
	\begin{tabular}{|l|c|c|c|c|}
		\hline
		\textbf{Model} &  \textbf{100k} & \textbf{1M} & \textbf{10M} & \textbf{100M} \\ \hline
		BM25 & 0.00\% & 0.00\% & 0.00\% & 0.00\% \\ \hline
		\multicolumn{5}{|l|}{Dense without hard negatives - MS MARCO} \\ \hline
		\quad 128 dim & 2.71\% & 4.41\% & 6.69\% & 9.73\% \\ \hline
		\quad 256 dim & 2.39\% & 4.03\% & 6.16\% & 9.04\% \\ \hline
		\quad 768 dim & 2.13\% & 3.72\% & 5.77\% & 8.52\% \\ \hline
		\multicolumn{5}{|l|}{Dense with hard negatives - MS MARCO} \\ \hline
		\quad 128 dim & 2.87\% & 4.20\% & 6.00\% & 8.11\% \\ \hline
		\quad 256 dim & 2.45\% & 3.72\% & 5.59\% & 7.38\% \\ \hline
		\quad 768 dim & 2.12\% & 3.32\% & 5.09\% & 7.03\% \\ \hline 
		\multicolumn{5}{|l|}{DPR \cite{dpr} - Natural Questions} \\ \hline
		\quad rnd title & 0.17\% & 0.28\% & 0.34\% & 0.51\% \\ \hline
		\quad all titles & 2.48\% & 5.59\% & 9.31\% & 12.08\% \\ \hline
		\quad dev titles & 4.18\% & 5.36\% & 6.66\% & 8.01\% \\ \hline
	\end{tabular}
	\caption{Percentage of queries for which a random string passage is ranked higher than the relevant passage. 100k/1M/10M/100M indicates the number of random passages in the index.}
	\label{table_rnd_noise}
\end{table}

We count for how many queries a random string is ranked higher than the relevant passage. The results are shown in Table \ref{table_rnd_noise}. We observe that BM25 does not rank any randomly generated passage higher than the relevant passage for the MS MARCO dataset. The chance that a random passage contains words matching the query is small.   

For the dense retrieval models, we observe for quite a large number of queries that a random string passage is ranked higher than the relevant passage. As proven in Section \ref{sec_theory}, the error increases with larger index sizes and fewer dimensions. 

For DPR, we observe an extreme dependency on the title. Having 100 million entries in the index with a real Wikipedia article title and a random paragraph, results in the retrieval of those for about 12.08\% of all questions at the top position.

The error numbers far exceed the estimation from equation (\ref{eq_false_pos}), confirming that the representations are not uniformly distributed over the complete vector space and are concentrated in a small space. In the appendix (Figure \ref{fig_umap}), we plot the representations for the queries, the relevant passages, and the random strings.

\section{Conclusion}
We have proven and shown empirically that the probability for false positives in dense information retrieval depends on the index size and on the dimensionality of the used representations. These approaches can even retrieve completely irrelevant, randomly generated passages with high probability. It is important to understand the limitations of dense retrieval:

1) Dense approaches work better for smaller, clean indexes. With increasing index size the difference to sparse approaches can decreases. 

2) Evaluation results with smaller indexes cannot be transferred to larger index sizes. A system that is state-of-the-art for an index of 1 million documents might perform badly on larger indices.

3) The false positive rate increases with fewer dimensions. 

4) The empirically found error rates far exceeded the mathematical lower-bound error rates, indicating that only a small fraction of the available vector space is effectively used.


\section*{Acknowledgments}
This work has been supported by the German Research Foundation through the German-Israeli Project Cooperation (DIP, grant DA 1600/1-1 and grant GU 798/17-1) and has been funded by the  German Federal Ministry of Education and Research and the Hessian Ministry of Higher Education, Research, Science and the Arts within their joint support of the National Research Center for Applied Cybersecurity ATHENE.

\bibliography{custom}
\bibliographystyle{acl_natbib}

\newpage
\appendix

\section{Plot of Random Noise Index}
\label{app_plot_rnd_noise_index}

\begin{figure*}[h]
	\centering
	\includegraphics[width=\linewidth]{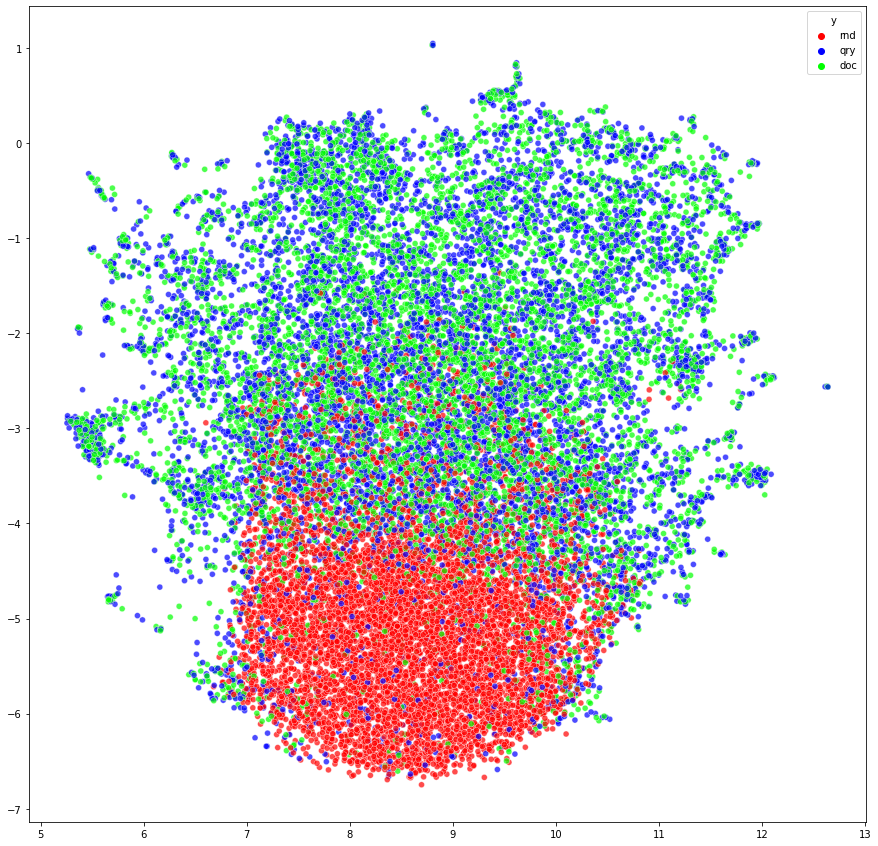}
	\caption{Plot of queries (blue), the relevant document (green) and representations from randomly generated strings (red). Dimensionality reduction via UMAP \cite{mcinnes2018umap-software}. Model with hard negatives, 768 dimensions. }
	\label{fig_umap}
\end{figure*}

Figure \ref{fig_umap} shows a two-dimensional plot of the 6980 development queries in the MS MARCO passage dataset, together with the 7433 passages that are marked as relevant and 7433 representations for randomly generated strings (using lowercase characters and space with a random length between 20 and 150 characters). The representation for the random strings are concentrated, but we still observe a significant overlap with the region for queries and relevant documents. This explains why random strings are retrieved for certain queries (Table \ref{table_rnd_noise}). We use the dense model that was trained with hard negatives with 768 dimensions. UMAP  \cite{mcinnes2018umap-software} is used for dimensionality reduction to 2 dimensions.

\end{document}